\documentclass[preprint,showpacs,preprintnumbers,amsmath,amssymb]{revtex4} 
 
\usepackage{graphicx}
\usepackage{dcolumn}
\usepackage{bm}
 
\begin{document} 

\title{Sum rules on quantum hadrodynamics}

\author{Bao-Xi Sun$^{2,\dagger}$, Xiao-Fu Lu$^{3,4,6}$, Peng-Nian Shen$^{1,2,3,6}$, 
En-Guang Zhao$^{1,3,5,6}$} 

\affiliation{${}^{1}$Center of Theoretical Nuclear Physics, 
National Laboratory of Heavy ion Accelerator,Lanzhou 730000,  China}
 
\affiliation{${}^{2}$Institute of High Energy Physics, The Chinese Academy of 
Sciences, P.O.Box 918(4), Beijing  100039, China} 
 
\affiliation{${}^{3}$Institute of Theoretical Physics, The Chinese Academy of 
Sciences, Beijing  100080, China} 
 
\affiliation{${}^{4}$Department of Physics, Sichuan University, 
Chengdu  610064, China} 
 
\affiliation{${}^{5}$Department of Physics, Tsinghua University, 
Beijing 100084, China} 
 
\affiliation{${}^{6}$China Center of Advanced Science and Technology(World 
Laboratory), Beijing 100080, China}

\begin{abstract}
The development on relativistic nuclear many-body theories is
reviewed. The second order self-energies of hadrons are calculated
from $\hat{S}_2$ matrix, and then an effective method to solve nuclear many-body
problems, sum rules on quantum hadrodynamics, is summarized.
The differences between this method and quantum hadrodynamics are discussed.
The effective nucleon mass in the nuclear matter is redefined in
the relativistic Hartree approximation, and a self-consistent relativistic mean-field model based on quantum hadrodynamics 2 is proposed.
It is pointed out that the original definition on effective nucleon mass in
quantum hadrodynamics 2 is not self-consistent, and all the parameters in the relativistic mean-field approximation should be fixed again according to the new definition.
\end{abstract}
\pacs{ 21.65.+f, 20.30.Fe }

\maketitle

\newpage

\section{Introduction}

\par

Although quantum field theory has succeeded greatly in the scope of
particle physics, people have met some difficulties when the
methods of the quantum field theory are applied to nuclear
systems. The ground state of the nuclear system is filled with
interacting nucleons,
which is different from the vacuum in perturbation theory. In
addition, The coupling constants of strong interaction between
nucleons are far larger than the fine structure constant in the
quantum electrodynamics, so the perturbation method could not
be performed perfectly on nuclear systems.
At last, the nucleons and mesons consist
of quarks and gluons, and are not fundamental components in the
level of modern knowledge. All these factors determine the
perturbation method in quantum field theory can only be
generalized to solve nuclear many-body problems approximately and
effectively.

Walecka and his group attempted to solve the nuclear many-body
problems in the framework of quantum field theory,
and developed the method of relativistic mean-field approximation
\cite{Wa.74,Chin.77,WS.86}. In this method, the field operators of
the scalar meson and vector meson are replaced with their
expectation values in the nuclear matter, respectively. Therefore,
the calculation is simplified largely.
Until now, there have been some excellent review articles giving a more
detailed description on this topic\cite{WS.86,Rein.89,Ring.96,WS.97}.
In the earliest relativistic mean-field theory, the resultant compression modulus is almost 550MeV\cite{Wa.74}, which is far from the experimental data
range of $200~-~300MeV$. To solve this problem, the nonlinear
self-coupling terms of scalar mesons are introduced to
produce the proper equation of state of nuclear
matter\cite{BB.77}. No doubt, additional parameters would
give more freedoms to fit the saturation curve of nuclear matter.
Zimanyi and Moszkowki developed the derivative scalar coupling
model yielding a compression modulus of 225MeV without any
additional parameter\cite{ZM.90}.

With the bare nucleon-nucleon interaction, the properties of
symmetric nuclear matter at various densities can be determined in
the relativistic Brueckner-Hartree-Fock calculation. For each
value of the density, the relativistic mean-field equations are
solved and the corresponding coupling constants are fixed to
the results of Brueckner calculation. Therefore, a relativistic
mean-field model with density-dependent coupling constants is
obtained\cite{Toki.92,Toki.97}.
Meanwhile, the Debye screening masses of mesons in the nuclear matter are
calculated in the relativistic mean-field
approximation\cite{Sun.0206}, and it shows all the screening meson masses increase with the nucleon number density. With
the meson masses replaced by their corresponding screening masses
in Walecka-1 model, the saturation properties of the nuclear
matter are fitted reasonably, and then a density-dependent
relativistic mean-field model is proposed.
The nonlinear self-coupling terms of the mesons are not
included in both of these density-dependent relativistic
mean-field models.
In the nuclear matter, the screening meson masses increase
with the density of the nuclear matter, and it is equivalent
to the statement that the coupling constants decrease with the
density increasing while the masses of mesons retain constant. At
this point, these two models are consistent with each other.

In the past decade, the relativistic mean-field theory has been
extended in various directions
to provide a more realistic description of nuclei.

Pair corrections in the nucleus are included
in the framework of relativistic Hartree-Fock-Bogoliubov
theory\cite{Ring.96,Ring.91}. Until now, this theory has been used
in the calculations of the properties of both the stable nucleus
and the nucleus far from $\beta$-stable
line\cite{Ring.96,Meng.98}.

Some attempts have been made to develop a model to solve nuclear
many-body problems in the quark level. The most representative
models are quark meson coupling model based on current
quarks\cite{Gu.88,Sa.97} and quark meson field 
model based on constituent quarks\cite{Toki.00,Tan.01}.

Quantum chromodynamics(QCD) has two very important properties
at low energies:
chiral symmetry spontaneous breaking and confinement. It is
believed that the two properties closely relate to the vacuum
characteristics of QCD. From the Goldstone's theorem, Goldstone
bosons appear as the chiral symmetry is spontaneously broken. To
carry out the constraint by appearance of Goldstone bosons,
several relativistic many-body models considering the chiral
symmetry spontaneous breaking have been developed
\cite{WS.97,Wein96,Pap99,LSZ.01}.

The relativistic mean-field results may be derived by
summing the tadpole diagrams self-consistently in nuclear matter,
retaining only the contributions from nucleons in the filled Fermi sea
in the evaluation of the self-energy and energy density, i.e., the relativistic mean-field method is
consistent to the relativistic Hartree approximation\cite{WS.86}. It is correct
only in the framework of original Walecka model(QHD-1). In the most popular QHD-2 model, in which the
nonlinear self-coupling terms of the scalar meson are included, whether the relativistic mean-field method is still consistent to the relativistic Hartree approximation has not be studied.
In Sect.~\ref{sect:wick}, we obtained the self-energies of hadrons
by calculating the second order $\hat{S}$ matrix 
in the nuclear matter, then an effective method to solve nuclear
many-body problems is evaluated. In Sect.~\ref{sect:scrha}, the effective nucleon mass in the nuclear matter is discussed with the nonlinear self-coupling terms of
the scalar meson included in the Lagrangian density, and then a
self-consistent relativistic mean-field scheme on quantum hadrodynamics 2 is proposed. The summary is given in Sect.~\ref{sect:sum}.

\section{The self-energies of hadrons in the nuclear matter}
\label{sect:wick}

According to Walecka-1 model, the nucleons $\psi$ interact with
scalar mesons $\sigma$ through a Yukawa coupling $\bar
\psi\psi\sigma$ and with neutral vector mesons $\omega$ that
couple to the conserved baryon current $\bar \psi \gamma_\mu
\psi$. the Lagrangian density can be written as
\begin{eqnarray}
\label{eq:Lagr}
 {\cal L}~&=&~\bar\psi
\left(i\gamma_{\mu}\partial^{\mu} -
M_N\right)\psi~+~\frac{1}{2}\partial_\mu\sigma\partial^\mu\sigma-\frac{1}{2}
m^2_\sigma \sigma^2_{} -\frac{1}{4}\omega_{\mu\nu}\omega^{\mu\nu}+
\frac{1}{2}
m^2_\omega\omega_\mu\omega^\mu \nonumber \\
&&-g_\sigma\bar\psi\sigma\psi-g_\omega\bar\psi \gamma_\mu
\omega^\mu \psi,
\end{eqnarray}
with $M_N$,$m_\sigma$ and $m_\omega$ the nucleon, scalar meson and
vector meson masses, respectively, and
$\omega_{\mu\nu}~=~\partial_\mu\omega_\nu-\partial_\nu\omega_\mu$
the vector meson field tensor.

The momentum-space propagators for the scalar meson, vector meson and the nucleon
take the forms of\cite{WS.86}
\begin{equation}
\label{eq:pro-scal}
 i{\Delta}(p) =
\frac{-1}{p^2-m^2_\sigma+i\varepsilon},
\end{equation}
\begin{equation}
\label{eq:pro-vect} i{D}_{\mu\nu}(p) =
\frac{g_{\mu\nu}}{p^2-m^2_\omega+i\varepsilon},
\end{equation}
\begin{equation}
\label{eq:GF}
 i{G_F}_{\alpha\beta}(p) = (\gamma_\mu
 p^\mu+M_N)_{\alpha\beta}\left( \frac{-1}{
  p^2-{M_N}^2+i\varepsilon}
\right).
\end{equation}
As the effect of Fermi sea is considered, an on-shell part 
\begin{equation}
\label{eq:GD}
 i{G_D}_{\alpha\beta}(p) = (\gamma_\mu
 p^\mu+M_N)_{\alpha\beta}\left( ~-~\frac{i\pi}{E(p)}
 \delta(p^0-E(p))\theta(p_F-|\vec{p}|)
\right),
\end{equation}
is included in the nucleon propagator besides the Feynman propagator in Eq.~(\ref{eq:GF}), 
where $E(p)=\sqrt{\vec{p}^2+M_N}$, and $p_F$ is the
Fermi momentum of nucleons.

Since the vector meson couples to the conserved baryon current, the
longitudinal part in the propagator of the vector meson will not contribute
to physical quantities\cite{IZ.80}. Therefore, only the transverse
part in the propagator of the vector meson is written
in Eq.~(\ref{eq:pro-vect}).

According to the Feynamn diagrams shown in Fig.1 and 2, the
self-energies of the nucleon, the scalar and vector meson in the
nuclear matter can be calculated with the Feynman rules in
Ref.~\cite{WS.86}. It should be noticed that there is a more
factor of $(i)$ in each of the propagators of the hadrons in our manuscript than those propagators in Ref.~\cite{WS.86}.

In this section, we will calculate the self-energies of hadrons
in the nuclear matter from Wick's theorem of quantum field theory.

The momentum-space
propagator of the nucleon takes the form of Feynman propagator
\cite{IZ.80}. Therefore, the Pauli blocking effect of Fermi sea is excluded
in the propagator of the nucleon, and the on-shell propagator is no use in the following calculations.

The interaction Hamiltonian can be expressed as
\begin{equation}
{\cal H}_I~=~ g_\sigma\bar\psi\sigma\psi+g_\omega\bar\psi
\gamma_\mu \omega^\mu \psi.
\end{equation}
In the second order approximation, only
\begin{equation}
\label{eq:S2S2}
\hat{S}_2 ~=~\frac{(-i)^2}{2 !}\int d^{4}x_1 \int d^{4}x_2
  T \left[{\cal H}_I(x_1){\cal H}_I(x_2) \right]
\end{equation}
in the S-matrix
should be calculated in order to obtain the self-energy corrections of the
nucleon and mesons.

\subsection{The nucleon self-energy in nuclear matter}
\label{sect:self}

The second order self-energy of the nucleon coupling to the scalar meson
is discussed firstly.

In order to obtain the second order self-energy correction of the nucleon
in Fermi sea, only the normal ordering product
\begin{equation}
\label{eq:0823}
 N\left[\bar\psi(x_1)\sigma\overbrace{(x_1)\psi(x_1)
\bar\psi(x_2)\sigma}(x_2)\psi(x_2) \right]
\end{equation}
in the Wick's expansion of the time-ordered product in Eq.~(\ref{eq:S2S2})
should be considered, where the overbrace "$\overbrace{}$"
denote the contraction of a pair of field operators.

When a nucleon with momentum $k$ and spin $\delta$ is considered
in the nuclear matter, its field operator $\psi(k, \delta, x)$ and
conjugate field operator $\bar\psi(k, \delta, x)$ can be expressed
as
\begin{equation}
\label{eq:pfield11}
 \psi(k, \delta, x) = A_{k\delta}U(k,\delta)\exp\left(-i k \cdot x\right)
+B^{\dagger}_{k\delta}V(k,\delta)\exp\left(i k \cdot x \right)
\end{equation}
and
\begin{equation}
\label{eq:bpfield11}
\bar\psi(k, \delta, x) =
A_{k\delta}^{\dagger}\bar{U}(k,\delta)\exp\left(ik \cdot
x\right)+B_{k\delta}\bar{V}(k,\delta)\exp\left(-i k \cdot
x\right),
\end{equation}
respectively.
In the calculation of the self-energy of the nucleon with the
momentum $k$ and the spin $\delta$, a pair of the nucleon field
operator and the conjugate operator in the normal ordering product
of Eq.~(\ref{eq:0823}) should be replaced with
Eqs.~(\ref{eq:pfield11}) and ~(\ref{eq:bpfield11}), while the
other pair of the nucleon field operator and the conjugate operator
connected with the underbrace
 "$\underbrace{}$" in the following normal ordering products denote
 the nucleon in the Fermi sea, and would be replaced with
their expansion forms of a complete set of solutions to the Dirac equation
, respectively.
\begin{eqnarray}
\label{eq:4th-term-1} &&
N\left[\bar\psi(x_1)\sigma\overbrace{(x_1)\psi(x_1)
\bar\psi(x_2)\sigma}(x_2)\psi(x_2) \right]  \\
&\rightarrow&2~\sigma\overbrace{(x_1)\sigma}(x_2)
\{N\left[\bar\psi(k, \delta,  x_1)\psi(k, \delta,  x_1)
\bar\psi\underbrace{(x_2)\psi}(x_2) \right] +
 N\left[\bar\psi(k, \delta,
x_1)\psi\underbrace{(x_1) \bar\psi}(x_2)\psi(k, \delta,  x_2)
\right] \}. \nonumber
\end{eqnarray}

Suppose there are no antinucleons in the ground state of nuclear
matter and the Fermi sea is filled with interacting nucleons, only
positive-energy components are considered in
the expansion forms of the nucleon field operator and its conjugation.
The expectation value of $\hat{S}_2$ in the nuclear matter on the
first term in Eq.~(\ref{eq:4th-term-1}) can be written as
\begin{eqnarray}
&&\label{eq:S2-1} \langle~G~|~\hat{S}_2~|~ G~\rangle
=ig_\sigma^2(2\pi)^4 \delta^4(p_1+k_1-p_2-k_2)
 \nonumber\\
&&\sum_{\lambda=1,2}\int\frac{d^3p}{(2\pi)^3}
\frac{M_N}{E(p)}\theta(p_F-|\vec{p}|)
\bar{U}(k,\delta)U(k,\delta)i\Delta(0)\bar{U}(p,\lambda)U(p,\lambda)
\nonumber 
\end{eqnarray}
where $k_1=k_2=k, $ and $p_1=p_2=p=(E(p),\vec{p}), $ and
$\theta(x)$ is the step function.

According to Dyson equation, the nucleon propagator 
in the nuclear matter can be derived as
\begin{eqnarray}
&&\frac{i}{\rlap{/}
k-M_N-\Sigma^\sigma_1+i\varepsilon}=\frac{i}{\rlap{/}
k-M_N+i\varepsilon}+ \frac{i}{\rlap{/}
k-M_N+i\varepsilon}\nonumber \\
&& i\frac{g_\sigma^2}{m_\sigma^2}
\sum_{\lambda=1,2}\int\frac{d^3p}{(2\pi)^3}
\frac{M_N}{E(p)}\theta(p_F-|\vec{p}|)
\bar{U}(p,\lambda)U(p,\lambda) \frac{i}{\rlap{/}
k-M_N+i\varepsilon},
\end{eqnarray}
then the second order self-energy of the nucleon in the nuclear matter from
the first term in Eq.~(\ref{eq:4th-term-1}) can be written as
\begin{eqnarray}
\label{eq:tadpole-self} \Sigma^\sigma_1
&=&-\frac{g_\sigma^2}{m_\sigma^2}
\sum_{\lambda=1,2}\int\frac{d^3p}{(2\pi)^3}
\frac{M_N}{E(p)}\theta(p_F-|\vec{p}|)
\bar{U}(p,\lambda)U(p,\lambda) \nonumber \\
&=&-\frac{g_\sigma^2}{m_\sigma^2}\rho_S
\end{eqnarray}
with
\begin{equation}
\label{eq:rhos}
\rho_S~=~\sum_{\lambda=1,2}\int\frac{d^3p}{(2\pi)^3}
\frac{M_N}{E(p)}\theta(p_F-|\vec{p}|)
\end{equation} 
the scalar density of protons or neutrons.

The second order self-energy of the nucleon relevant to the second term in
Eq.~(\ref{eq:4th-term-1}) can be obtained similarly
\begin{eqnarray}
\Sigma^{\sigma}_{2}&=&g^2_\sigma \sum_{\lambda=1,2}\int
\frac{d^3p}{(2\pi)^3}
\frac{M_N}{E(p)}\theta(p_F-|\vec{p}|)
\left[U(p,\lambda)~i\Delta(k-p)~\bar{U}(p,\lambda)
\right]  \\
&=&-g^2_\sigma \int \frac{d^3p}{(2\pi)^3}
\frac{M_N}{E(p)}\theta(p_F-|\vec{p}|)
\left[\frac{\rlap{/} p+M_N}{2M_N}
\frac{1}{(k-p)^2-m^2_\sigma}\right]. \nonumber
\end{eqnarray}

Correspondingly, the normal ordering products relevant to
the self-energy of the nucleon coupling to the vector meson in the
calculation of the $\hat{S}_2$ matrix can be written as
\begin{eqnarray}
\label{eq:4th-term-vector} &&
N\left[\bar\psi(x_1)\gamma_\mu\omega^\mu\overbrace{(x_1)\psi(x_1)
\bar\psi(x_2)\gamma_\nu\omega^\nu}(x_2)\psi(x_2) \right]  \\
&\rightarrow&2~\omega^\mu\overbrace{(x_1)\omega^\nu}(x_2)
\nonumber \\
 &&\{N\left[\bar\psi(k, \delta, x_1)\gamma_\mu\psi(k, \delta, x_1)
\bar\psi\underbrace{(x_2)\gamma_\nu\psi}(x_2) \right] +
 N\left[\bar\psi(k, \delta,
x_1)\gamma_\mu\psi\underbrace{(x_1)
\bar\psi}(x_2)\gamma_\nu\psi(k, \delta, x_2) \right] \}. \nonumber
\end{eqnarray}
Therefore, the second order self-energies of the nucleon coupling
to the vector
meson in the nuclear matter corresponding to the first and second
terms in Eq.~(\ref{eq:4th-term-vector}) can be calculated as
\begin{eqnarray}
\label{eq:tadpole-omega-1}
\Sigma_{1}^{\omega}&=&(-ig_\omega)^2 \sum_{\lambda=1,2}\int
\frac{d^3p}{(2\pi)^3} \frac{M_N}{E(p)} \theta(p_F-|\vec{p}|)\nonumber \\
&& \gamma_\mu~iD^{\mu\nu}(0) \left[\bar{U}(p,\lambda)\gamma_\nu
U(p,\lambda)
\right] \nonumber \\
&=&\gamma_0 \frac{g^2_\omega}{m^2_\omega} \rho_V
\end{eqnarray}
with
\begin{equation}
\label{eq:rhov}
\rho_V=\sum_{\lambda=1,2}\int\frac{d^3p}{(2\pi)^3}
\theta(p_F-|\vec{p}|) \end{equation}
 the number density of protons or neutrons, and
\begin{eqnarray}
\Sigma^{\omega}_{2}&=&g^2_\omega \sum_{\lambda=1,2}\int
\frac{d^3p}{(2\pi)^3}
\frac{M_N}{E(p)}\theta(p_F-|\vec{p}|) \left[ \gamma_\mu
U(p,\lambda)~iD_0^{\mu\nu}(k-p)~\bar{U}(p,\lambda) \gamma_\nu
\right] \nonumber \\
&=&g_\omega^2 \int
\frac{d^3p}{(2\pi)^3}\frac{1}{(\vec{p}^2+{M_N}^2)^\frac{1}{2}}
\frac{-\gamma_\mu p^\mu+2M_N}{(k-p)^2-m^2_\omega}.
\end{eqnarray}

Obviously, the second order self-energies of the nucleon
$\Sigma^{\sigma}_{1}$,
$\Sigma^{\sigma}_{2}$, $\Sigma^{\omega}_{1}$,
$\Sigma^{\omega}_{2}$ calculated from Wick's expansion in this
section are same as those results from quantum hadrodynamics, respectively\cite{WS.86}.

\subsection{The self-energies of the scalar and vector mesons}
\label{sect:sig-ome}

In the Wick expansion of the time-ordered
product in Eq.~(\ref{eq:S2S2}), only
the normal ordering products including one contraction of a pair
of nucleon field operator and its conjugate operator should be studied
in order to obtain the second order self-energy of the scalar
 meson in the filled Fermi sea.

When a scalar meson with determined momentum $k$ is studied in the
nuclear matter, its field operator $\sigma(k,x)$ can be expressed
as
\begin{equation}
\label{eq:sigma-k} \sigma(k,x)~=~ a_{k}\exp(-ik \cdot x)
+a^\dagger_{k}\exp(ik \cdot x).
\end{equation}
In order to calculate the self-energy of the scalar meson in the
nuclear matter, the scalar field operators $\sigma(x_1)$ and
$\sigma(x_2)$ in the following normal ordering product
should be replaced with
Eq.~(\ref{eq:sigma-k}),
\begin{eqnarray}
\label{eq:sigma-rele}
&&N\left[\bar\psi\overbrace{(x_1)\sigma(x_1)\psi(x_1)
\bar\psi(x_2)\sigma(x_2)\psi}(x_2) \right]
+N\left[\bar\psi(x_1)\sigma(x_1)\psi\overbrace{(x_1)
\bar\psi}(x_2)\sigma(x_2)\psi(x_2) \right]\nonumber \\
&\rightarrow& 2\psi\overbrace{(x_1)
\bar\psi}(x_2)N\left[\bar\psi\underbrace{(x_1)\sigma(k,x_1)\sigma(k,x_2)\psi}(x_2)\right].
\end{eqnarray}
The nucleon field operator $\psi(x_2)$ and the conjugate field
operator $\bar \psi(x_1)$ in the normal ordering product
of Eq.~(\ref{eq:sigma-rele}) should be expanded in terms of the set of
solutions to the Dirac equation, respectively.
Therefore, the second order self-energy of the scalar meson can be obtained
as\cite{Sun.0206}
\begin{eqnarray}
\label{eq:scalar-self} \Sigma_\sigma&=&
(-ig_\sigma)^2~\sum_{\lambda=1,2} \int\frac{d^3p}{(2\pi)^3}
\frac{M_N}{E(p)}\theta(p_F-|\vec{p}|) \nonumber \\
&&\left[ \bar{U}(p,\lambda) \left(iG(p-k)+iG(p+k)\right)
 U(p,\lambda)\right] \nonumber \\
&=&g^2_\sigma
\int\frac{d^3p}{(2\pi)^3}\frac{M_N}{E(p)}\theta(p_F-|\vec{p}|)
\left[ Tr\left( \frac{1}{\rlap{/} p-\rlap{/} k-M_N}
\frac {\rlap{/} p+M_N}{2M_N} \right) \right. \nonumber \\
&&+\left. Tr\left(\frac{\rlap{/} p+M_N}{2M_N}
\frac{1}{\rlap{/} p+\rlap{/} k-M_N} \right)
 \right].
\end{eqnarray}

Similarly, the normal ordering products relevant to the
self-energy of the vector meson in the nuclear matter can be written as
\begin{eqnarray}
&&N\left[\bar\psi\overbrace{(x_1)\gamma_\mu
\omega^\mu(x_1)\psi(x_1) \bar\psi(x_2)\gamma_\nu
\omega^\nu(x_2)\psi}(x_2) \right] \nonumber \\
&+&N\left[\bar\psi(x_1)\gamma_\mu
\omega^\mu(x_1)\psi\overbrace{(x_1)
\bar\psi}(x_2)\gamma_\nu \omega^\nu(x_2)\psi(x_2) \right]\nonumber \\
&\rightarrow& 2\psi\overbrace{(x_1)
\bar\psi}(x_2)N\left[\bar\psi\underbrace{(x_1)\gamma_\mu
\omega^\mu(k,\delta,x_1)\gamma_\nu
\omega^\nu(k,\delta,x_2)\psi}(x_2)\right]
\end{eqnarray}
with
\begin{equation}
\omega_\mu(k,\delta,x)~=~b_{k
\delta}\varepsilon_\mu(k,\delta)\exp(-ik \cdot x)
+b^\dagger_{k\delta}\varepsilon_\mu(k,\delta)\exp(ik \cdot x).
\end{equation}

The self-energy of the vector meson in the nuclear matter can be
calculated similarly as
\begin{eqnarray}
\label{eq:vector-self} \Sigma_\omega&=&
(-ig_\omega)^2~\sum_{\lambda=1,2}\int\frac{d^3p}{(2\pi)^3}
\frac{M_N}{E(p)}\theta(p_F-|\vec{p}|) \nonumber \\
&&\left[\bar{U}(p,\lambda)
\left(\gamma_{\nu}iG(p-k)\gamma_{\mu}~+~ \gamma_{\mu}iG(p+k)
\gamma_{\nu}\right) U(p,\lambda) \right] \nonumber \\
 &=&g^2_\omega
\int\frac{d^3p}{(2\pi)^3}\frac{M_N}{E(p)}\theta(p_F-|\vec{p}|)
\left[ Tr\left(\gamma_\nu \frac{1}{\rlap{/} p-\rlap{/} k-M_N}
\gamma_\mu \frac{\rlap{/} p+M_N}{2M_N}
 \right)
 \right. \nonumber \\
&& + \left. Tr\left(\gamma_\nu \frac{\rlap{/}
p+M_N}{2M_N} \gamma_\mu \frac{1}{\rlap{/} p+\rlap{/}
k-M_N} \right)
 \right].
\end{eqnarray}

It is no doubt that the second order
self-energies of the scalar and vector mesons calculated from
$\hat{S}_2$ matrix directly are same as those from quantum
hadrodynamics, respectively\cite{WS.86}.

\subsection{Feynman rules}

The second-order
self-energies of the nucleon, the scalar and vector meson in the
nuclear matter are calculated from Wick's expansion. It shows the same results as those in quantum hadrodynamics\cite{WS.86}, and then an effective many-body method based on vacuum propagators has been evaluated. Feynman rules on this effective method can be summarized similarly as those in quantum hadrodynamics\cite{WS.86}. In the new Feynman rules, a factor of  
$$\sum_{\lambda=1,2}\int\frac{d^3p}{(2\pi)^3}
\frac{M_N}{E(p)}\theta(p_F-|\vec{p}|)$$ is included for each pair of
crosses, which denote the initial and final states of the nucleon in the Fermi sea. Moreover, the momentums and spins of external lines with a cross or
without a cross take the same values with each other,
respectively. In addition, include a factor of $(-1)$ in the calculation of exchange
diagrams.

The loop diagrams, which relate to the contribution of Dirac sea and cause
divergences, are not
necessary to be considered in no Dirac sea approximation.
Therefore, only the diagrams with
crosses should be studied in the calculation of self-energies of
particles. Because there do not exist antinucleons in the ground
state of nuclear matter, the diagrams with an external line of
antinucleons should be excluded, too.

The Feynman diagrams for the second order self-energy of the nucleon
in the nuclear matter in Section
~\ref{sect:self} are shown in Fig. 3. The first diagram in Fig.3
corresponds to the tadpole contribution, and the second
corresponds to the exchange term in the relativistic Hartree-Fock
approximation\cite{WS.86}.

The self-energies of the scalar meson, vector meson or the photon in the nuclear
matter can be calculated with the Feynman diagrams in Fig.4. It shows the same results as the one-fermion-loop approximation in quantum
hadrodynamics\cite{WS.86}, i.e., the same effective
masses of the photon and mesons in Ref.\cite{Sun.0206,Sun.02}
can be obtained in the one-fermion-loop approximation in quantum hadrodynamics.

In our formalism, the effects of the nuclear medium come from the nucleon condensation, i.e., the scalar density of nucleons. In Walecka's formalism, the propagators of hadrons are defined in the ground state of the nuclear matter and different from the propagators defined in vacuum, and the loop diagrams are considered although the meanings are different from those in quantum field theory.
Therefore, Feynman rules in our formalism are different from those in
Ref.~\cite{WS.86,Furn.91}, and The condensation of the nucleon is embodied in the integrals of three-momentum space in the new Feynman rules.

In particle physics, people are mostly interested in scattering processes, for which the $\hat{S}$ matrix providing the probability of transition from the initial states to final states, is the most suitable framework. In statistical physics, however, we are mainly concerned on the expectation value of physical quantities at finite time. Obviously, these two problems are connected with each other in our formalism.
Because vacuum propagators are adopted in our formalism, which are not relevant to the state of the system, it is not difficult to extend this formalism to study the properties of non-equilibrium and finite temperature states. Some works have been done along this direction\cite{Sun.0209}.

Actually, the propagator including the on-shell part of
Eq.~(\ref{eq:GD}) is not for the nucleon, but for a kind of quasinucleon, whose creation and annihilation operators satisfy the same
anticommutation relations as those of the nucleon. There is a Bogoliubov
transformation between the creation and
annihilation operators of the quasinucleon and the nucleon\cite{Ume.82}.

\section{Self-consistent relativistic Hartree approximation}
\label{sect:scrha}

When the isospin $SU(2)$ symmetry is considered in the nuclear matter, the
$\rho$ meson interaction should be included in the Lagrangian density,
\begin{equation}
{\cal L}^{\rho}_{Int}= -~g_{\rho}  \bar\psi \gamma^\mu \frac{\vec{\tau}}{2}
\cdot \vec{\rho}_\mu \psi
\end{equation}
with $\vec{\tau}$ being the Pauli matrix.
Because the $\rho_{+}$ and $\rho_{-}$ mesons only contribute to the second
order self-energy of the nucleon in the exchange terms, only the
$\rho_{0}$ meson interaction is considered in the relativistic Hartree
approximation.

With the similar method, the second order
self-energy corrections of the proton and neutron coupling to the
$\rho_{0}$ meson
can be written as
\begin{equation}
\label{eq:p-rho}
\Sigma_{1}^{\rho}~=~\gamma_0 \frac{g^2_\rho}{\pm 4 m^2_\rho}
\left(\rho_p - \rho_n \right)
\end{equation}
with the plus for the proton and the minus for the neutron,
where $\rho_p$ and $\rho_n$ are the number density of protons and neutrons,
respectively. Obviously, the results in Eq.~(\ref{eq:p-rho})
are same as those in the relativistic mean-field approximation\cite{WS.86}.

In the calculation of the second order self-energies of hadrons in the
nuclear matter in Sect.~\ref{sect:wick},
the noninteracting propagators of hardrons are used. Although the
second order results can be summed to all orders with Dyson's equation,
this procedure is not self-consistent. Self-consistency can be achieved
by using the interacting propagators to also determine the
self-energy\cite{WS.86}. In the relativistic Hartree
approximation, the self-energy of the nucleon in the nuclear matter
can be calculated self-consistently with the interacting propagator
of the nucleon
\begin{equation}
\label{eq:GF-2}
 i{G_H}(p) = \frac{-1}{\gamma_\mu
 \bar p^\mu-{M^\ast_N}+i\varepsilon},
\end{equation}
where
\begin{equation}
\label{eq:effn}
M^\ast_N=M_N-\frac{g_\sigma^2}{m_\sigma^2}\left(\rho^S_p
+ \rho^S_n\right),
\end{equation}
\begin{equation}
\bar p^0=p^0~-~\frac{g^2_\omega}{m^2_\omega} \left( \rho_p + \rho_n \right)
~-~\frac{g^2_\rho}{\pm 4 m^2_\rho}
\left(\rho_p - \rho_n \right)
\end{equation}
with the plus for the proton and the minus for the neutron, and
\begin{equation}
\vec{\bar p}=\vec{p}.
\end{equation}
In Eq.~(\ref{eq:effn}), $\rho^S_p$ and $\rho^S_n$ are the scalar densities
of protons and
neutrons, respectively. It corresponds to the transformation
\begin{equation}
M_N\rightarrow M^\ast_N,~~~~~~E(p)\rightarrow E^\ast(p),
\end{equation}
in the self-energy of the nucleon in Eq.~(\ref{eq:tadpole-self})
and~(\ref{eq:tadpole-omega-1}),
where $E^\ast(p)=(\vec{p}^2+{M^\ast_N}^2)^{1/2}$.

The effective nucleon mass can be defined
as the pole of the nucleon propagator in the limit of the
space-momentum of the nucleon $\vec{p}\rightarrow 0$, which corresponds
to the mass spectra of the collective excitations
in the nuclear matter\cite{Song.93,SRK.95}.
According to Eq.~(\ref{eq:tadpole-self}), the effective nucleon mass in the nuclear matter is defined in Eq.~(\ref{eq:effn}) in the relativistic Hartree approximation.
In quantum hadrodynamics 2, the nonlinear
self-coupling terms of the scalar meson are introduced to
replace the mass term $\frac{1}{2} m^2_\sigma \sigma^2_{}$
\cite{BB.77},
\begin{equation}
U(\sigma)~=~\frac{1}{2} m^2_\sigma \sigma^2_{} ~+~\frac{1}{3} g_2
\sigma^3_{}~+~\frac{1}{4} g_3 \sigma^4_{}.
\end{equation}
Because the boson distribution functions of the mesons are zero in the
nuclear matter at zero temperature, the self-coupling terms of the scalar
meson have no contribution to the self-energy corrections of the nucleon and the
meson when the loop diagrams are ignored. Therefore, the effective nucleon 
mass still takes the form in Eq.~(\ref{eq:effn})
in quantum hadrodynamics 2, which is important to conserve the
self-consistency in the
calculation of relativistic Hartree approximation.

In the relativistic mean-field approximation of quantum hadrodynamics 2,
\begin{equation}
m^2_\sigma \sigma_0~+~g_2 \sigma^2_{0}~+~g_3 \sigma^3_{0}
=-{g_\sigma}(\rho^S_p+\rho^S_n),
\end{equation}
then the effective nucleon mass in the nuclear matter can be
written as
\begin{equation}
\label{eq:effnrmf2}
M^{\ast}_N~=~M_N~+~g_\sigma \sigma_0
~+~\frac{g_{\sigma} g_2}{m^2_{\sigma}} \sigma^2_{0}
~+~\frac{g_{\sigma} g_3}{m^2_{\sigma}} \sigma^3_{0}.
\end{equation}

The effective nucleon mass in the relativistic mean-field approximation has not been studied carefully although the other four saturation properties of the nuclear matter
have been given more attentions. From the relativistic mean-field form of Dirac Equation, the
{\sl effective} nucleon mass, which is called Dirac mass in our manuscript,
had been drawn out by adding the $g_\sigma  \sigma_0$ on the mass of the
nucleon, which had been believed to be
correct In QHD-1 and QHD-2. From this definition, The {\sl effective} nucleon mass is only
related to the linear term of the scalar meson.  In QHD-1, it is same as
the effective nucleon mass derived from the relativistic Hartree approximation in Eq.~(\ref{eq:effn}). In the relativistic Hartree approximation of QHD-2, The
effective nucleon mass still takes the form of Eq.~(\ref{eq:effn}). If the
self-consistency of the relativistic mean-field approximation of QHD-2 is considered, which is very important in the calculation of strong interaction systems, the effective nucleon mass
must be defined as Eq.~(\ref{eq:effn}), then the nonlinear self-coupling terms of the scalar meson will contribute to the effective nucleon mass
directly in the relativistic mean-field approximation. Moreover, the form of the effective nucleon mass in Eq.~(\ref{eq:effnrmf2})is consistent with
the definition in Ref.~\cite{Bethe.71}.

In the quantum hadrodynamics 2, the total energy density and the
pressure of nuclear matter can be deduced to
\begin{equation}
\varepsilon = \frac{1}{2} {m}^2_\sigma \sigma^2_{0} + \frac{1}{3}
{g}_2 \sigma^3_{0} + \frac{1}{4} {g}_3 \sigma^4_{0} +\frac{1}{2}
m^2_\omega \omega^2_{0} + \frac{1}{2}m^2_\rho \rho^2_{0} +
\sum_{B=p,n} \frac{2}{\left( 2\pi \right)^3} \int^{p_{F}(B)}_{0}
d\vec{p} \left(\vec{p}^2 + {M_{N}^{\dagger}}^2
\right)^{\frac{1}{2}},
\end{equation}
and
\begin{eqnarray}
p~&=&~\frac{1}{3}\sum_{B=p,n} \left(
\frac{2}{\left( 2\pi \right)^3}
\int^{p_{F}(B)}_{0} d\vec{p} \left(\vec{p}^2 +
{M_{N}^{\dagger}}^2 \right)^{\frac{1}{2}}
 \right) \nonumber \\
&& - \frac{1}{2} {m}^2_\sigma \sigma^2_{0}
- \frac{1}{3} {g}_2 \sigma^3_{0} - \frac{1}{4} {g}_3 \sigma^4_{0}
 + \frac{1}{2} {m}^2_\omega \omega^2_{0} + \frac{1}{2} {m}^2_\rho
 \rho^2_{0},
\end{eqnarray}
where $M_{N}^{\dagger}~=~M_N~+~g_\sigma \sigma_0$ is
the Dirac mass of the nucleon in the nuclear matter
defined from the Dirac equation of the nucleon.

Because the original definition of the effective nucleon mass is wrong, all sets of parameters on the relativistic mean-field approximation of QHD-2 based on that wrong definition can not be used in the new self-consistent scheme mentioned above. Therefore, all the parameters must be readjusted.
By fitting the saturation properties of nuclear matter, the
parameters in the relativistic mean-field approximation with the effective nucleon 
mass defined in Eq.~(\ref{eq:effnrmf2})
 can be fixed as
\begin{eqnarray}
&&g_\sigma~=~8.95,~~g_\omega~=~10.94,~~g_\rho~=~7.2, \nonumber  \\
&&g_2~=~-1.38fm^{-1},~~ g_3~=~23.0~~ \nonumber
\end{eqnarray}
with $m_\sigma~=~532.5MeV$, and the masses of the vector and $\rho$ mesons
take the experimental values, respectively.

With these parameters  we obtain a saturation density of
$0.1655\mbox{fm}^{-3}$, a binding energy of $15.771$~MeV, a
compression modulus of $227$~MeV, an asymmetric energy coefficient
of $32.44$~MeV and an effective nucleon mass of $0.638M_N$ for the
symmetric saturation nuclear matter. At the saturation point,
the Dirac mass of
the nucleon is $0.661M_N$, larger than the effective nucleon mass
under this set of parameters. More accurate parameters should be fixed by
studying the properties of finite nuclei.

In Walecka's model, the effective nucleon mass is defined from the Dirac equation of the nucleon in the relativistic mean-field approximation. However, the effective nucleon mass is defined as the pole of the nucleon propagator in the relativistic Hartree approximation in our model. Although these two definitions are same as each other in the framework of $QHD-1$, they are different from each other in $QHD-2$, in which the self-coupling terms of the scalar meson are included. Since the self-consistency is realized in a different manner, the parameters must be fixed again to fit the saturation properties of nuclear matter.

As far as the self-consistency of calculations for strong interaction systems is concerned, the redefinition on the effective nucleon mass is essential.
The original definition of effective nucleon mass in quantum hadrodynamics is wrong and the relativistic mean-field calculations with parameters fixed according to this wrong definition are not self-consistent.

Because of the strong coupling between hadrons, the nuclear systems can only be studied approximately and effectively in the framework of quantum field theory, and
the renormalization is even meaningless on nuclear systems. In nuclear many-body theories, the self-consistency should be considered, and the different interacting propagators should be adopted in the calculation of different order approaches.

\section{Summary}
\label{sect:sum}

In a conclusion, an effective formalism to solve nuclear many-body
problems is evaluated, and we find this formalism
with off-shell propagators gives the same results as
those in quantum hadrodynamics
in the calculation of
self-energies of particles in the nuclear matter. Moreover,
Feynman rules is induced for this new method, which is named as sum
rules on quantum hadrodynamics by us. In addition, the self-consistency of quantum hadrodynamics 2 is discussed in the relativistic mean-field approximation.

${}^\dagger$ Corresponding author,
E-mail address: $sunbx@mail.ihep.ac.cn$.
This work was supported in part by the Major State Basic Research
Development Program under Contract No. G2000-0774, the CAS
Knowledge Innovation Project No. KJCX2-N11 and the National
Natural Science Foundation of China No. 10075057, 90103020 and
10047001.

\newpage

\leftline{\Large {\bf Figure Captions}}
\parindent = 2 true cm
\parskip 1 cm
\begin{description}

\item[Fig. 1] \label{fig:rhfa-loop} Feynman diagrams for
the second order self-energy of the nucleon in
nuclear matter calculated in the quantum hadrodynamics.
The double solid lines denote the nucleon propagators defined 
in the ground state of nuclear matter.\\

\par

\item[Fig. 2] \label{fig:meson-loop} Feynman diagram for the second order
self-energies of the scalar or vector meson in
nuclear matter calculated in the quantum hadrodynamics.
Same case as in Fig. 1.\\

\par

\item[Fig. 3]  Feynman diagrams for the second order self-energy of the
nucleon in
nuclear matter calculated from $\hat{S}_2$ matrix elements. The
wave lines denote the scalar meson or vector meson, while 1 and 2
denote particles of the initial
state, 3 and 4 denote particles of the final state.\\

\par

\item[Fig. 4]  Feynman diagrams for the second order self-energies of
the scalar or vector meson in
nuclear matter calculated from $\hat{S}_2$ matrix elements. Same case as
in Fig. 3. \\

\end{description}

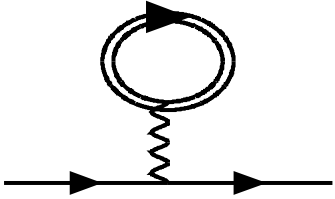
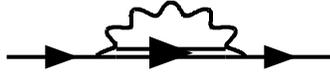
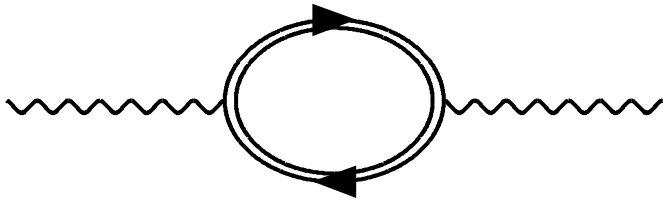
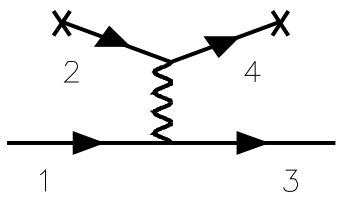
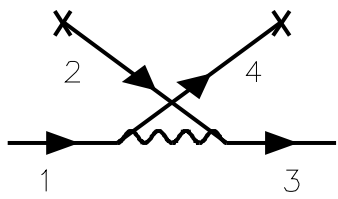
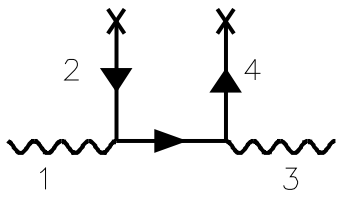
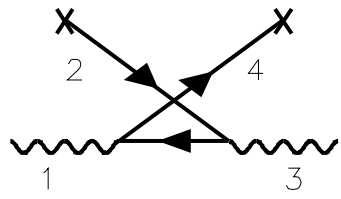
\end{document}